\documentclass[aps,amssymb,amsmath, onecolumn, pra, superscriptaddress]{revtex4}
\usepackage{graphicx}
\usepackage{epsfig}
\usepackage{dcolumn}
\usepackage{amsmath}
\usepackage{bm}
\usepackage{bbm}
\usepackage[colorlinks, citecolor=red]{hyperref}
\usepackage{color}
\usepackage{soul}
\usepackage{centernot}
\usepackage[up]{subfigure}
\usepackage[normalem]{ulem}	

\DeclareMathOperator{\Tr}{Tr}

\newcommand{\be}{\begin{equation}}
\newcommand{\ee}{\end{equation}}
\newcommand{\bc}{\begin{center}}
\newcommand{\ec}{\end{center}}
\newcommand{\bea}{\begin{eqnarray}}
\newcommand{\eea}{\end{eqnarray}}
\newcommand{\ba}{\begin{array}}
\newcommand{\ea}{\end{array}}

\begin{document}
\title{ Quantum percolation and transition point of a directed discrete-time quantum walk}
\author{C. M. Chandrashekar}
\email{c.madaiah@oist.jp}
\author{Th. Busch}
\affiliation{Quantum Systems Unit, Okinawa Institute of Science and Technology Graduate University, Okinawa, Japan}

\begin{abstract}
\noindent
 Quantum percolation describes the problem of a quantum particle moving through a disordered system. While certain similarities to classical percolation exist, the quantum case has additional complexity due to the possibility of Anderson localisation. Here, we consider a directed discrete-time quantum walk as a model to study quantum percolation of a two-state particle on a two-dimensional lattice. Using numerical analysis we determine the fraction of connected edges required (transition point) in the lattice for the two-state particle to percolate with finite (non-zero) probability for  three fundamental lattice geometries, finite  square lattice,  honeycomb lattice, and nanotube structure and show that it tends towards unity for increasing lattice sizes. To support the numerical results we also use a continuum approximation to analytically derive the expression for the percolation probability for the case of the square lattice and show that it agrees with the numerically obtained results for the discrete case.  Beyond the fundamental interest to understand the dynamics of a two-state particle on a lattice (network) with disconnected vertices, our study has the potential to shed light on the transport dynamics in various quantum condensed matter systems and the construction of quantum information processing and communication protocols. 
 \end{abstract}
\maketitle
Percolation theory, which describes the dynamics of particles in random media\,\cite{K73, BR06}, is an established area of research with numerous applications in diverse fields\,\cite{SS94, KE09}. The main figure of merit which quantifies the transport efficiency of a particle in percolation theory is the so-called percolation threshold\,\cite{SA94}. To illustrate its meaning in the classical setting, one can consider transport on a square lattice with neighbouring vertices connected with probability $p$. For $p=0$, all vertices are disconnected from each other and no path for the particle to move across the lattice exists. With increasing $p$ more and more vertices will be connected and once $p=p_c = 0.5$  a connection across the full lattice is established. 

The corresponding problem of percolation of a quantum particle differs from the classical setting in that quantum interference plays a significant role\,\cite{O86, MDS94, LN09}. One consequence of that is that in a random or disordered system the interference of the different phases accumulated by the quantum particle along different routes during the evolution can lead to the particle's wave function becoming exponentially localized. This process is well known as Anderson localisation\,\cite{And58, LR85, EM08} and has recently been experimentally observed in different disordered systems\,\cite{SBF07, CLG08, COR13}.
Using a one parameter scaling argument it has been shown that all two-dimensional (2D) Anderson systems are exponentially localised for any amount of disorder\,\cite{AALR79}. Quantum interference therefore becomes as important in quantum percolation as the existence of the connection between the vertices, making it a more intriguing setting when compared to the classical counterpart\,\cite{KE72, SAB82, VW92, SF09}. 

Transport of a two-state quantum system (qubit) across a large network is an important process in quantum information processing and communication protocols\,\cite{NC00} and by today many physical systems are tested for their scalability and engineering properties.  Furthermore, in last couple of years quantum transport models have also shown a certain applicability to understanding transport processes in biological and chemical systems\,\cite{ECR07, MRL08, PH08}. These natural or synthetic systems are not guaranteed to have a perfectly connected lattice structure and can possess a directed evolution in any particular direction. Therefore, it is important to consider the possible role quantum percolation can play in understanding transport in these systems and the effects of directed evolution on the Anderson localization length (the spatial spread of the localized state). In this work we present a model which is discrete in time to study the percolation of two-state quantum particle on a two-dimensional lattice with directed transport in one of the dimensions. This study will complement the previously reported studies on quantum percolation using continuous-time Hamiltonian (see articles in lecture notes\,\cite{LN09} and references therein).

To model the dynamics of the two-state quantum particle we choose the process of quantum walks\,\cite{Ria58, Fey86, Par88, ADZ93, Mey96}, which in recent years has been shown to be an important and highly versatile mechanism\,\cite{Sal12}. Recently, first studies of two-state quantum walks in percolating graphs have been reported for circular and linear geometries\,\cite{KKN12} as well as for square lattices using a four-state particle\,\cite{OPD06, LKB10}. Here we present the physically applicable model of a directed discrete-time quantum walk (D-DQW) of a two-state particle to study quantum percolation on a 2D lattice. Not unexpectedly we find a non-zero percolation probability on a lattice of finite size when the fraction of missing edges is small. An increase in this fraction, however, quickly results in a zero percolation probability highlighting the importance of well-connected lattice structure for quantum percolation. Using the continuum approximation of the discrete dynamics we then derive an analytical expression for the percolation probability and show that it is in perfect agreement with the numerical result. Since the percolation threshold tends towards unity for large lattices, we find that the directed evolution in also agreement with the scaling prediction of localization for any amount of disorder\,\cite{AALR79}.

Below we will first establish a benchmark by describing the dynamics of  the D-DQW on a completely connected two-dimensional lattice and introduce the modifications necessary to describe the dynamics when some of the connections between the vertices are missing. In {\bf Results} we present the numerically obtained percolation probabilities for different fractions of disconnections between the vertices and for different lattice geometries and also give the analytically obtained formula for the square lattice case in the continuum limit. All of these results are interpreted in {\bf Discussion} and we finally detail the analytical approach in {\bf Methods}.
\vskip 0.2in
\noindent
{\bf Directed discrete-time quantum walk ~:~}
Let us first define the dynamics of a D-DQW on a completely connected square lattice of dimension $n \times n$.  The Hilbert space of the complete system is given by  ${\cal H} = {\cal H}_c \otimes {\cal H}_l$, where the space of the particle (coin space) ${\cal H}_c$ contains its internal states, $|\downarrow \rangle = \begin{bmatrix} 1  \\ 0 \end{bmatrix}$ and $|\uparrow \rangle =\begin{bmatrix} 0  \\ 1 \end{bmatrix}$ and the space of the square lattice ${\cal H}_l$ contains the vertices $(x, y)$ of the lattice, $|x, y\rangle$. For each step of the D-DQW we will consider the standard DQW evolution in the $x$ direction\,\cite{ADZ93, Mey96}, followed by the directed evolution in the $y$ direction, which is based on a scheme presented by Hoyer and Meyer\,\cite{HM09}.  The evolution in $x$ direction on a completely connected lattice consists of the coin-flip operation
 \begin{align}
\label{eq:1}
C_\theta  \equiv &  \begin{bmatrix} \begin{array}{clcr}
  \mbox{~~~}\cos(\theta)      &     &   -i \sin(\theta)
  \\ -i \sin(\theta) & &  \mbox{~~~}\cos(\theta) \end{array} \end{bmatrix},
  \end{align}
 followed by the shift along the connected edges 
 \begin{align}
\label{eq:2}
S_x^c\equiv & \sum_{x}\sum_{y } \Big ( |\downarrow \rangle\langle\downarrow|\otimes |x-1, y \rangle\langle x, y |   + | \uparrow \rangle\langle\uparrow|\otimes  |x+1, y \rangle\langle x, y|  \Big ). 
   \end{align}
Here $|\downarrow\rangle$ and $|\uparrow \rangle$ can equivalently be used to indicate the edge along the negative or positive $x$ direction. For the directed evolution along the positive $y$ direction we define the walk using one directed edge and $r-1$ self-looping edges at each vertex  $(x, y)$ and assign a basis vector to each edge \cite{HM09}. Thus, every state at each edge is a linear combination of the states
\begin{align}
| +\rangle \otimes |x, y\rangle , | \circlearrowleft_1 \rangle \otimes |x, y\rangle, | \circlearrowleft_2 \rangle \otimes |x, y\rangle,\cdots \cdots, | \circlearrowleft_{r-1} \rangle \otimes |x, y\rangle,
\end{align}
where $|+\rangle$ indicates the edge along the positive $y$ direction  and each $|\circlearrowleft\rangle$ indicates a distinct self-loop. As shown in Ref.\,\cite{HM09}, this is equivalent to an effective coin space $\{+, \circlearrowleft \}$  at each vertex and the coin-flip operation can be defined as
\begin{align} 
\label{eq:cy}
C_y  \equiv &  \begin{bmatrix} \begin{array}{clcr}
  \alpha      &     &   ~~~\beta
  \\  \beta & &  -\alpha \end{array} \end{bmatrix}, ~~\mbox{where}~~\beta = \sqrt{\frac{r-1}{r}} ~~~, ~~~\alpha = \frac{1}{\sqrt r },
  \end{align}
with the shift along the edges  given by
  \begin{align}
\label{eq:2y}
S_y^c\equiv & \sum_{x} \sum_{y } \Big ( | + \rangle\langle + |\otimes |x, y+1 \rangle\langle x, y |   + | \circlearrowleft \rangle \langle \circlearrowleft|\otimes  |x, y \rangle\langle x, y|  \Big ). 
   \end{align}
The state of the two-state walker at any vertex is therefore given by $|\psi_{x, y}\rangle = (\alpha_{x, y} |\downarrow \rangle  + \beta_{x,y} |\uparrow \rangle) \otimes |x, y\rangle = \psi^{\downarrow}_{x, y} + \psi^{\uparrow}_{x, y} \equiv  \psi^{+}_{x, y} + \psi^{\circlearrowleft}_{x, y}$, where the latter identity indicates that the edge dependent basis states for $y$ direction can be the same as the ones for the $x$ direction. However, in general the operation corresponding to one complete step of the D-DQW on a completely connected lattice will then be
\begin{align}
W^c = \Big [S_{y}^c \big( C_y \otimes \mathbb{I}_x  \otimes \mathbb{I}_y \big ) \Big]_{+, \circlearrowleft}  \Big [S_{x}^c\big (C_{\theta} \otimes \mathbb{I}_x  \otimes \mathbb{I}_y \big) \Big ]_{\downarrow, \uparrow},
\end{align}
where the subscripts on the square brackets represent the basis on which the operators act. 
\bc 
\begin{figure}[th!]
\includegraphics[width=9.5cm]{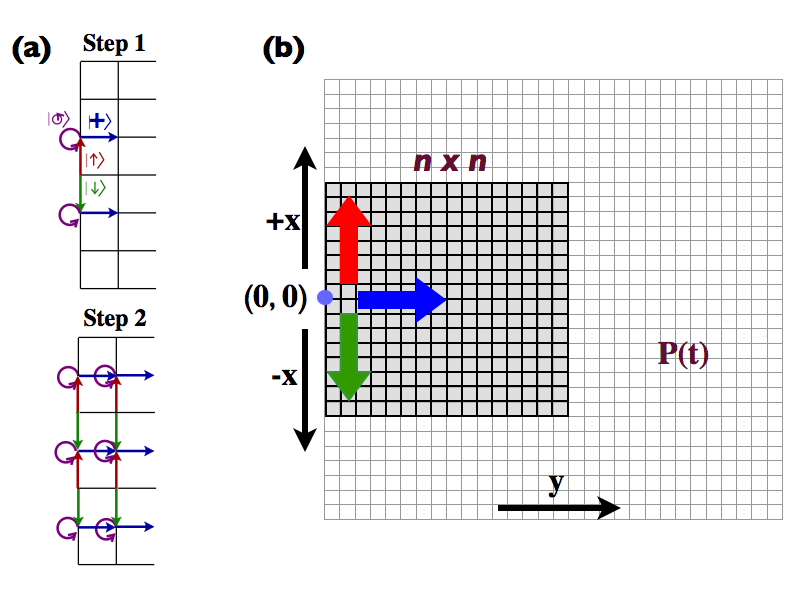}  
\vskip 0.0cm
\caption{\footnotesize{{\bf Schematic describing the D-DQW on a square lattice.} (a) Schematic of the first two steps of the D-DQW of a two-state particle with states $|\downarrow \rangle$ and $|\uparrow \rangle$, representing the edges along the negative and positive $x$ direction, and $|+\rangle$ and $|\circlearrowleft\rangle$, representing the directed edge along the positive $y$ direction and the self-looping edges at each vertex. (b) Schematic of the direction of the spread of the wavepacket on the lattice.  $P(t)$ is the probability of finding the particle outside of the sub-lattice lattice of dimension $n \times n$ after time $t$, which corresponds to the percolation probability of the D-DQW.} \label{fig:1}}
\end{figure} 
\ec
In Fig.\,\ref{fig:1}(a) we show a schematic for the first two steps on a two-dimensional lattice. With the particle initially at the origin, $|\Psi_{in}\rangle = \left ( \cos(\delta/2)| \downarrow \rangle + e^{i\eta} \sin(\delta/2)|\uparrow \rangle \right )\otimes | 0, 0\rangle$, the state  after $t$ steps is given by
\be
 |\Psi_t\rangle =[W^c]^t |\Psi_{in} \rangle, 
\ee
and in Fig.\,\ref{fig:1}(b) the direction of the spread of the wavepacket on the lattice is indicated. To find the probability of detecting the particle outside of a sub-lattice of size $n\times n$ on an infinitely large lattice after $t$ steps one then has to calculate
\be
P(t ) =1 - \Tr \Bigg( \sum_{x = -\lfloor  n/2\rceil}^{~\lfloor  n/2\rceil} ~~\sum_{y =0}^{n-1} \langle x, y | \rho(t)  |x, y\rangle \Bigg),
\ee
where $\rho(t) = |\Psi_t \rangle \langle \Psi_t |$ and the term $\Tr(\cdots)$  describes the probability of finding the particle on the sub-lattice.  For a one-dimensional DQW the probability distributions are well known to spread over the range  $[- t \cos(\theta),  t \cos(\theta)]$ and to decrease exponentially outside that region\,\cite{NV01, CSL08}, and for a one-dimensional D-DQW  the interval of spread is given by $[ \frac{1-\alpha}{2} t, \frac{1+\alpha}{2}t]$\,\cite{BP07}. 
\bc
\begin{figure}[th!]
 \includegraphics[width=13.0cm]{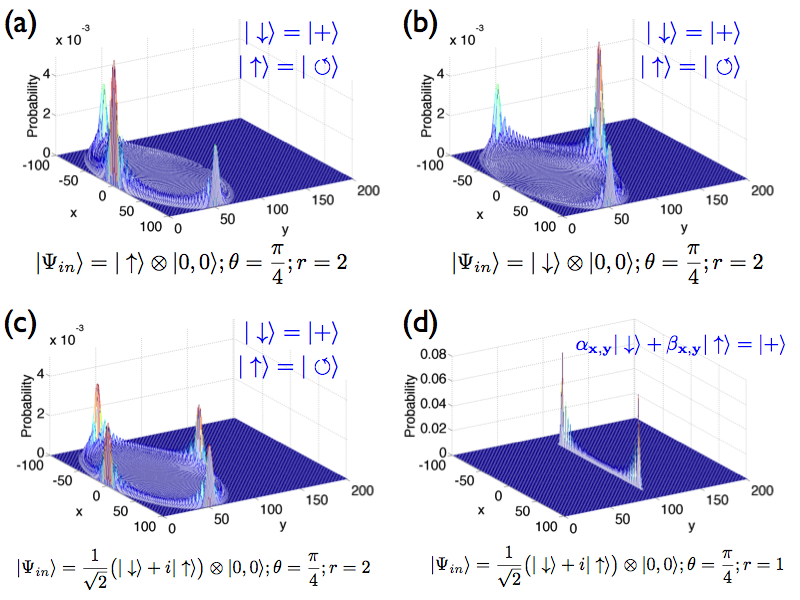}  
\caption{\footnotesize{{\bf Probability distributions of the D-DQW on a lattice of size $200\times 200$ after 100 steps of walk.}  Probability distribution after 100 steps with different initial state of the particle,  {\bf (a)}  $|\uparrow\rangle$, {\bf (b)} $|\downarrow\rangle$ and {\bf (c)} $\frac{1}{\sqrt 2} \big(|\downarrow \rangle + i |\uparrow \rangle \big )$ with the same basis states along the edges in both, $x$ and $y$ direction, and using the values of $\theta = \pi/4$ and $r =2$ in the coin operation. Each distribution is spread over the interval $[-t, t]$ along $x$ direction and $[0, t]$ along $y$ direction. The ones resulting from the initial states $|\uparrow\rangle$ and $|\downarrow\rangle$ are asymmetric along the $y$ direction and mirror each other with respect to $y=t/2$, whereas for the initial state $\frac{1}{\sqrt 2} \Big(|\downarrow \rangle + i |\uparrow \rangle \Big )$ the distribution is symmetric with respect to $x=0$ and $y=t/2$. {\bf (d)} For the basis state $|+\rangle = \alpha_{x, y} |\downarrow \rangle + \beta_{x, y} |\uparrow \rangle$, the dynamical model results in a 1D DQW along $x$ direction and directed movement in $y$ direction.} \label{fig:2}}
\end{figure} 
\ec

For the two-dimensional D-DQW described above we show the probability distributions in Fig.\,\ref{fig:2} for $t=100$, $\theta = \pi/4$, $r=2$ and the same basis states along the $x$ and $y$ direction ($|\downarrow \rangle = |+\rangle$ and $|\uparrow \rangle =| \circlearrowleft\rangle$), and one can clearly see that the spread in the $x$ direction is over the interval $[-t, t]$ and in the $y$ direction across the range $[0,t]$. The details of the evolved probability distributions depend of the initial state of the particle (see Figs.\,\ref{fig:2}(a), (b) and (c)) and for long time evolutions ($t \rightarrow \infty$) the probability for the particle to be found outside of a finite sub-lattice will approach $P(t) \rightarrow 1$. 

\begin{figure}[tb]
\bc 
\includegraphics[width=8.8cm]{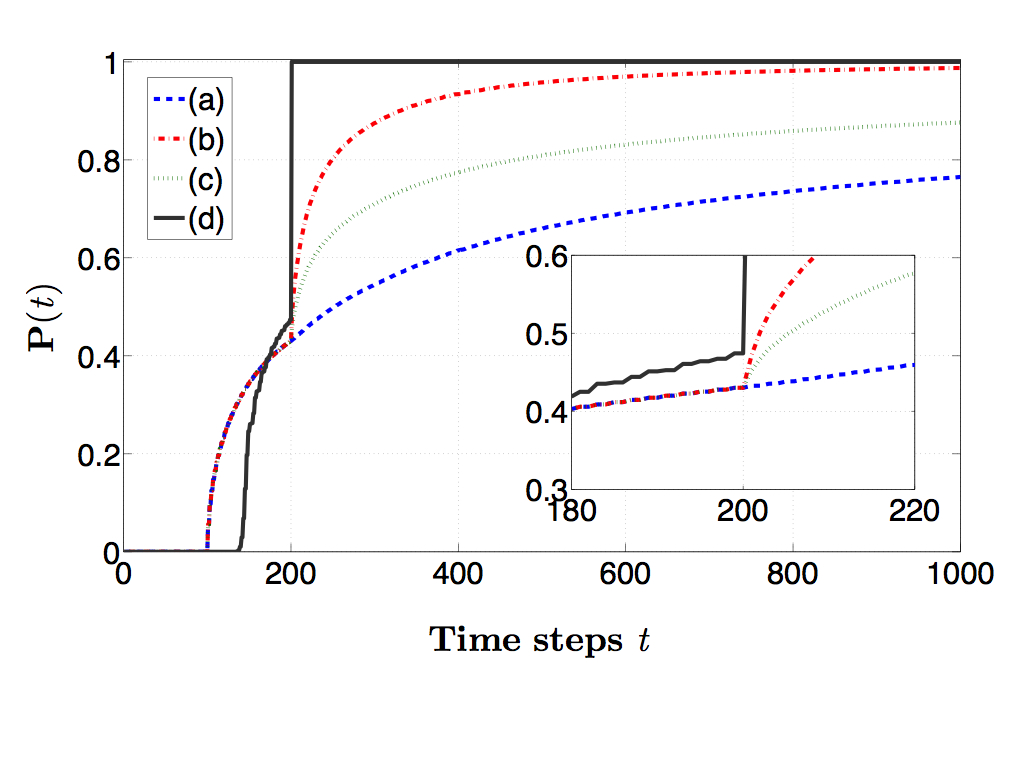}  
\ec
\caption{\footnotesize{{\bf Probability $P(t)$ of finding the particle outside a sub-lattice of size $200 \times 200$ as a function of time steps $t$.} For evolution with the basis states $|\downarrow \rangle = |+\rangle$, $|\uparrow \rangle = |\circlearrowleft \rangle$  when $\theta =\frac{\pi}{4}$, $r=2$ and the initial state of the particle is $|\uparrow \rangle$, $|\downarrow \rangle$ or  $\frac{1}{\sqrt 2} \big(|\downarrow \rangle + i |\uparrow \rangle \big)$ (corresponding to plots (a) to (c) in Fig.\,\ref{fig:2}). Irrespective of the initial state,  the increase of $P(t)$ is the same for $t\le n$, after which the asymmetry in the probability distribution leads to differences.  For $\theta =\frac{\pi}{4}$ and $r=1$,  with $|+\rangle = \alpha_{x, y} |\downarrow \rangle + \beta_{x, y} |\uparrow \rangle$ (corresponding to plot (d) in Fig.\,\ref{fig:2}), the probability can be seen to jump to $P(t) = 1$ for $t > n$, as expected from the directed movement in the $y$ direction. The inset shows the details of the behaviour around $t= 200$.}\label{fig:3}}
\end{figure} 
In case of a missing self-looping edge, the basis state along the edge connecting $y$ with $y+1$ will be specific to the vertex, $|+\rangle = \alpha_{x, y} |\downarrow \rangle + \beta_{x, y} |\uparrow \rangle$. This corresponds to a 1D DQW traversing a 2D lattice, as shown in Fig.\,\ref{fig:2}(d), and complete transfer ($P(t) =1$) will already be achieved for $t > n$.  In Fig.\,\ref{fig:3} we show the probability of finding the particle outside the sub-lattice of size $200 \times 200$ as a function of time steps $t$ for all the four cases of Fig.\,\ref{fig:2}. Choosing a different value of $\theta$ in the coin operation along the $x$ direction would lead to different rates of increase in $P(t)$.

\vskip 0.2in
\noindent
{\bf Directed discrete-time quantum walk on a lattice with missing edges~:~}
Let us now consider lattice structures in which some of the edges connecting the vertices are missing. For the evolution along the $x$ direction, for which the basis states are $|\downarrow \rangle$ and $|\uparrow \rangle$, the coin operation will be $C_{\theta}$\,[Eq.\,(\ref{eq:1})] and the shift operator has to be defined according to the number of edges at each vertex. If both edges in the $x$-direction are present we have
\begin{align}
S_{\forall \,x \rightarrow x^{\prime}} \equiv &   |\downarrow \rangle\langle\downarrow|\otimes |x-1, y \rangle\langle x, y |  + | \uparrow \rangle\langle\uparrow|\otimes  |x+1, y \rangle\langle x, y|,
   \end{align}
   where $x^{\prime}$ is  $(x+1)$ and $(x-1)$, whereas the absence of even one of the edges requires
\begin{align}
S_{\forall \,x \centernot \rightarrow x^{\prime}}\equiv &   |\downarrow \rangle\langle\downarrow|\otimes |x, y \rangle\langle x, y | 
    + | \uparrow \rangle\langle\uparrow|\otimes  |x, y \rangle\langle x, y|.
   \end{align}
Alternatively, this shift operator can be written using a different self-looping edge for both basis states. The operator corresponding to each step along the $x$ direction is then given by 
\begin{align}
S_{x} (x, y) \big( C_{\theta} \otimes \mathbb{I}_x  \otimes \mathbb{I}_y \big)=\big(S_{\forall \,x \rightarrow x^{\prime}} + S_{\forall x \centernot \rightarrow x^{\prime}} \big ) \big( C_{\theta} \otimes \mathbb{I}_x  \otimes \mathbb{I}_y \big).
\end{align}
A similar description applies to the evolution in the $y$ direction. When an edge connecting $(x, y)$ and $(x, y+1)$ is present, the shift operator  will be
\begin{align}
S_{\forall  \,y \rightarrow y^{\prime}} \equiv &   |+ \rangle\langle + |\otimes |x, y+1 \rangle\langle x, y | 
    + | \circlearrowleft \rangle\langle \circlearrowleft |\otimes  |x, y \rangle\langle x, y|,
\end{align}
where $y^{\prime}$ is $(y+1)$.  When this edge is missing, both  states, $|+\rangle$ and $|\circlearrowleft\rangle$ are basis states for different self-looping edges and the shift operator will be
\begin{align}
S_{\forall  \,y \centernot \rightarrow  y^{\prime}} \equiv &   |+ \rangle\langle + |\otimes |x, y \rangle\langle x, y | 
    + | \circlearrowleft \rangle\langle \circlearrowleft |\otimes  |x, y \rangle\langle x, y|.
  \end{align}
 Thus, the operator corresponding to evolution along the $y$ direction can be written as
 \begin{align}
S_{y} (x, y) \big( C_{y} \otimes \mathbb{I}_x  \otimes \mathbb{I}_y \big)=
\big(S_{\forall \,y \rightarrow y^{\prime}} + S_{\forall y \centernot \rightarrow y^{\prime}} \big ) \big( C_{y} \otimes \mathbb{I}_x  \otimes \mathbb{I}_y \big),
\end{align}
and one complete step of the D-DQW on a lattice which is not fully connected is given by
\begin{align}
W_d = \Big [S_{y}(x, y) \big( C_y \otimes \mathbb{I}_x  \otimes \mathbb{I}_y \big ) \Big]_{+, \circlearrowleft}  \Big [S_{x} (x, y)\big (C_{\theta} \otimes \mathbb{I}_x  \otimes \mathbb{I}_y \big) \Big ]_{\downarrow, \uparrow}.
\end{align}
Here again the subscripts on the square brackets indicate the basis states of the edges for the evolution along each direction and the state of the particle after $t$ steps is then given by
\be
 |\Psi_t^d\rangle =[W_d]^t |\Psi_{in} \rangle.
 \ee

\begin{figure}[th!] 
\includegraphics[width=15.0cm]{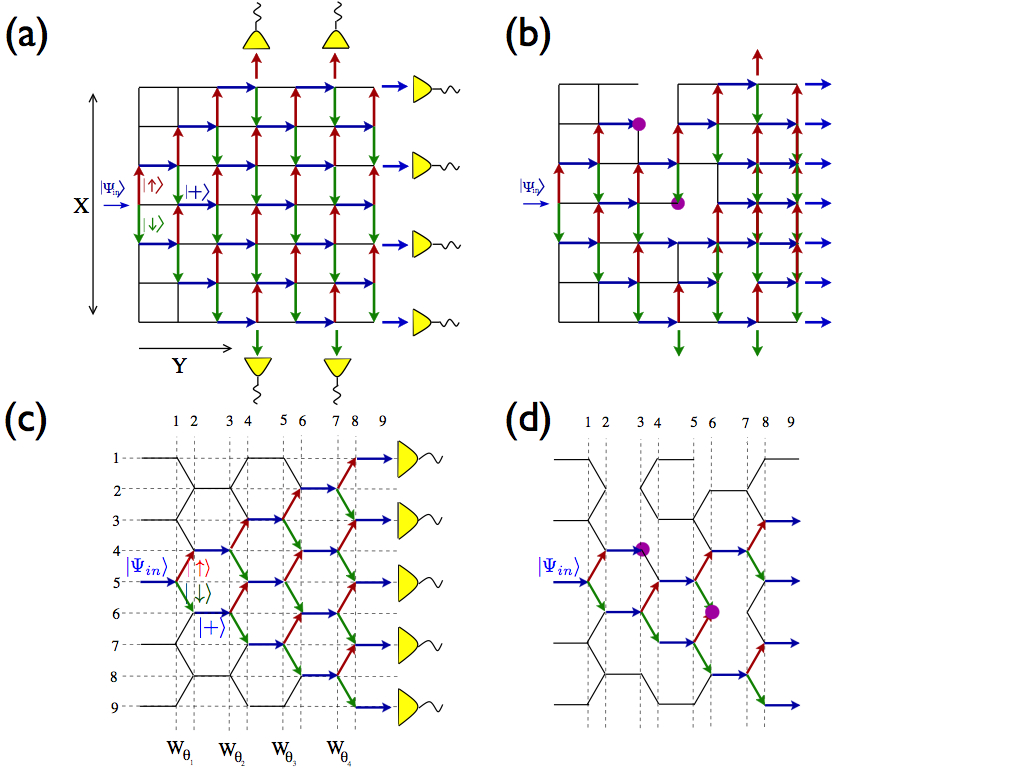} 
\caption{\footnotesize{{\bf Schematic of the paths possible for a two-state particle on exemplary square and honeycomb lattices.} Green, red and blue arrows represent the direction of the shift for $|\downarrow \rangle$, $|\uparrow \rangle$ and both the states, respectively. {\bf (a)} and {\bf (c)} show the possible paths when all vertices are perfectly connected and {\bf (b)} and {\bf (d)} show the paths when some connections are missing. The positions at which localization occurs are highlighted.} \label{fig:4}}
\end{figure}

In a classical setting the percolation threshold for a square lattice can be calculated to be $p_c=0.5$ and is known to be independent of the lattice size. In a quantum system, however, a disconnected vertex breaks the ordered interference of the multiple traversing paths, which can result in the trapping of a fraction of the amplitude at this point. This disturbance of the interference due to the disorder is known to result in Anderson localization \cite{And58, LR85} and consequently a large percentage of connected vertices is required to reach a non-zero probability for the particle to cross the lattice. In the following we will define the percolation probability as 
\begin{align}
\zeta(p) =1 - \Tr \left( \sum_{-\lfloor  n/2\rceil}^{~\lfloor  n/2\rceil} ~~\sum_{y =0}^{n-1} \langle x, y | \rho^d(t , p)  |x, y\rangle \right),
\end{align}
where $\rho^d(t, p) = |\Psi_t \rangle \langle \Psi_t |$ and  $\Tr (\cdots )$ is the probability of finding the particle in the sub-lattice of size $n \times n$ as  $t\rightarrow \infty$ for a fixed percentage of disconnected vertices $p$.  From this description of the dynamics we can note that a state gets trapped at a vertex $(x, y)$ 
only if the edge connecting $(x, y)$ with $(x, y+1)$ and one or both  of the edges connecting $(x, y)$ with $(x+1, y)$ and $(x-1, y)$ are missing. From all other vertices a state eventually finds a path and moves on. Therefore, the probability of finding a particle trapped on the sublattice for $t\rightarrow \infty$ will stem only from states trapped at vertices with a missing edge along the positive $y$ direction and one or two missing edges along $x$ direction. We will call the critical value at which the fraction of connected vertices $p$ is large enough to reach a certain, non-zero $\zeta(p)$ the  transition point, $p_a$. While the value chosen for this is, of course, somewhat arbitrary, the results presented below only require our choice to be consistent and we therefore  fix $\zeta(p) = 0.01$. In fact, for any values smaller than $0.01$, no noticeable changes are seen in our simulations. The transition point is then obtained by averaging over a large number of realisations, in which an initial state is evolved for a time $t$ that is large enough to give all parts of the state a chance to find their way through the lattice and contribute to the probability. To make this numerically efficient we will in the following concentrate on the case where the basis state along $y$ is given by $|+\rangle = \alpha_{x, y}|\downarrow \rangle + \beta_{x, y}|\uparrow \rangle$, as it has the narrowest probability distribution in the $y$ direction for a perfectly connected lattice. A schematic of the walk for a fully connected square lattice is shown in Fig.\,\ref{fig:4}(a) and an example for a broken lattice in Fig.\,\ref{fig:4}(b). 
\bc
\begin{figure}[th!] 
\includegraphics[width=11.0cm]{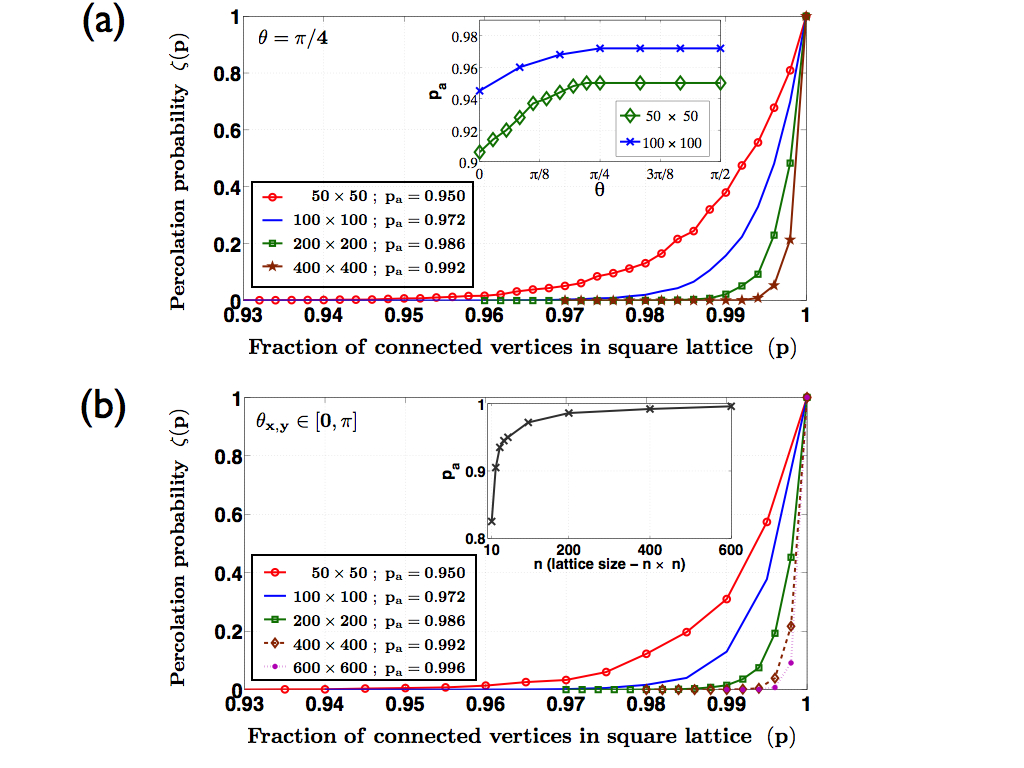} 
\caption{\footnotesize{{\bf Percolation probability as funding of percentage of connections for square lattice.} We can see an increase in the percolation probability as a function of the  percentage of connections for square lattices of different sizes using a coin with (a) $\theta = \pi/4$ and (b) $\theta_{x, y} \in [0, \pi]$.   The values for the transition points are shown as a function of $\theta$ and lattice size in the inset of {\bf (a)} and {\bf (b)}, respectively. They can be seen to approach unity for increasing lattice size and also show a strong dependence on $\theta$ for $\theta < \pi/4$ and the lattice size for small values. } \label{fig:5}}
\end{figure}
\ec
\par
In addition to considering quantum percolation on the fundamental square lattices, transport processes on honeycomb lattices and nanotubes have attracted considerable attention in recent years \cite{SAH11} and the two-state quantum percolation model can be expected to give useful insight into the behaviour of quantum currents and their transition points. In Fig.~\ref{fig:4}(c) we show the path taken by the two-state D-DQW on a honeycomb lattice of dimension $9 \times 9$ and in Fig.~\ref{fig:4}(d) we give an example for the situation where some connections are missing. For the later case each step of the D-DQW consists of 
\begin{align}
W_d^H = \Big [S_{y}(x, y) W_y \Big]_{+, \circlearrowleft}  \Big [S^{\prime}(q)W_{\theta} \Big ]_{\downarrow, \uparrow},
\end{align}
where the quantum coin operators, $W_{\theta}=C_{\theta} \otimes \mathbb{I}_x  \otimes \mathbb{I}_y$ and 
$W_y = C_y \otimes \mathbb{I}_x  \otimes \mathbb{I}_y$, and the shift operator for the directed evolution in the $y$-direction, $S_{y}(x, y)$, are same ones as the ones used for the evolution on the square lattice.  Due to the honeycomb geometry however, the shift operator $S^{\prime}(q)$ transition $W_q$ from $q$ to $q \pm 1$ corresponds to a shift along two edges, first in the $\pm x$ direction and then along the positive $y$ direction.

 \vskip 0.2in
\bc
{\bf \Large Results}
\ec
{\bf Numerical~:~}
In Fig.~\ref{fig:5}(a) we show the average $\zeta(p)$  for square lattices of different sizes, which have been obtained by taking the arithmetic mean of 200 independent numerical realizations and ensuring that the error bars are small. One can immediately see that $p_a$ is significantly larger than $p_c=0.5$, and even grows towards unity with increasing  lattice size, which is a clear indicator of the important role Anderson localisation plays in the dynamical process. The inset of Fig.~\ref{fig:5}(a) shows the dependence of $p_a$ on $\theta$ and the increase visible for larger angles can be understood by first considering a square lattice with all vertices fully connected and the initial position given by $(x, y) = (0, 0)$. If the coin parameter is chosen as $\theta = 0$, the two basis states move away from each other along the $x$-axis (no interference takes place) and exit from the sides at $t=n/2$.  For finite values of $\theta$, the exit point is pushed towards the positive $y$ direction due to interference in the $x$ direction and on an imperfect lattice the walker encounters potentially more broken connections (which lead to additional interferences). For small values of $\theta$ however, a large fraction of the state will still exit along the sides without interference and therefore only a smaller number of connections  are needed.  

\begin{figure}[th!]
\bc 
\includegraphics[width=9.0cm]{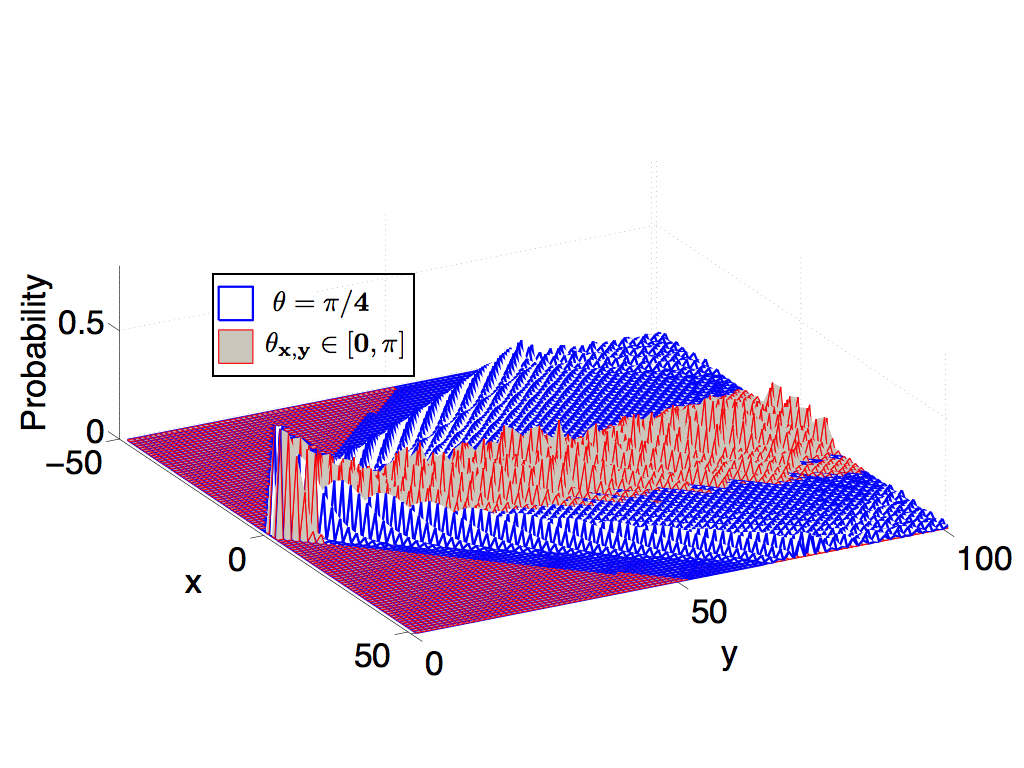}
\ec
\vskip -0.9cm
\caption{\footnotesize{{\bf Spread of the probability distribution in the $x$-direction for the D-DQW with $\theta_{x, y}$ (red) and $\theta = \pi/4$ (blue).}  The size of the completely connected lattice is $100 \times 100$. The red data shows a single representation using a position dependent coin, $\theta_{x,y} \in [0, \pi]$, and a much smaller spread in the $x$ direction is clearly visible. }\label{fig:6}}
\end{figure}
\par
The assumption of having the same coin operator at each lattice site is a rather strong one and in the following we will relax this condition to account for applications in more realistic situations. For this we replace $\theta$ by a vertex dependent parameter, $\theta_{x, y} \in [0 , \pi]$ that does not only account for local variations, but also allows for $\cos(\theta_{x,y})$ to be negative if $\theta_{x,y} \in [\pi/2, \pi ]$. This corresponds to a displacement of the left moving component to the right and of the right moving component to the left in the $x$-direction, which in turn can lead to localization in transverse direction\,\cite{Cha12}. To illustrate the effect of this, an example for a single realisation is shown in Fig.~\ref{fig:6} for a completely connected square lattice. A significantly reduced transversal spread compared to a walk with $\theta=\pi/4$ is clearly visible.
The average percolation probability as a function of $p$ for this evolution is shown in Fig.\,\ref{fig:5}(b) and, interestingly, we find that the disorder in the form of $\theta_{x,y}$ does not result in any noticeable change in the value of $p_a$ when compared to $\theta=\pi/4$.  This is due to that fact that $\theta_{x, y}$ and $\theta = \pi/4$ give rise to nearly the same degree of interference \cite{Cha12} and it highlights the dominance of the localization effects along the transverse direction. 
\par
\bc
\begin{figure}[th!] 
\includegraphics[width=11.0cm]{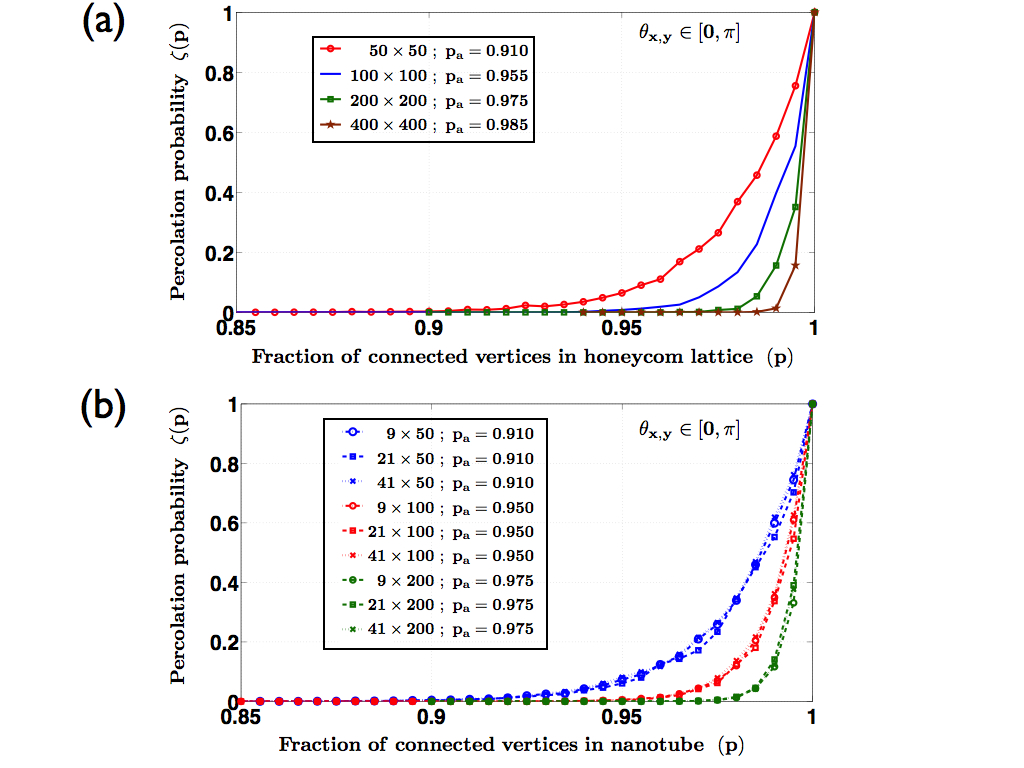} 
\caption{\footnotesize{{\bf Percolation probability for honeycomb lattice and nanotube geometry.} We can see an increase in percolation probability and  transition point as function of the percentage of connections for a honeycomb lattice and a nanotube geometry. Percolation probability as a function of percentage of connected vertices for (a) honeycomb lattices and (b) nanotube structures of different sizes. For the particle transport process the value of $\theta$ has been randomly picked from $[ 0 , \pi]$  at each vertex. With increase in lattice size, $p_a$ shifts towards unity and for nanotubes of size $n \times y$ it can be seen to be independent of $n$.}\label{fig:7}}
\end{figure}
\ec
The resulting average percolation probability for an honeycomb lattice is shown in Fig.\,\ref{fig:7}(a) as a function of the percentage of connected vertices with randomly assigned value of $\theta_{x,y} \in [0,\pi]$. Again, the data points were obtained by taking the arithmetic mean of 200 independent realizations and ensuring that the error bars a small. Similarly to the case  of the square lattice we find that $p_a$ is significantly larger than the classical percolation threshold $p_c=0.652$ \cite{SE64} and also lattice size dependent.  Note that, compared to a square lattice of the same size, $p_a$ for a honeycomb lattice is smaller, which gives the honeycomb structure an advantage over the square lattice for quantum percolation using a two-state D-DQW. This can be understood by considering the geometry of the honeycomb lattice: the edges in the honeycomb lattice are such that the both operators $S^{\prime}(q)$ and $S_y(x, y)$  contribute to the shift in $y$ direction, whereas in the square lattice only $S_y(x, y)$ contributes to this shift. 

An interesting extension to the honeycomb lattice is the introduction of periodic boundary conditions in the $x$-direction, 
which transforms the flat lattice into a nanotube geometry. This
corresponds to allowing transitions from $q$ to $(q \pm 1 \text{ mod }n)$, where $n$ is the number of vertices along the $x$-axis and in Fig.~\ref{fig:7}(b) we show the percolation probability for such a structure as a function of the percentage of connected vertices with randomly assigned values of $\theta_{x,y} \in [0, \pi]$ at each vertex. One can see that the
 transition point is same as that of a flat honeycomb structure with the same number of edges in the transverse direction, which can be understood by realising that the periodic boundary conditions increase the probability for a particle to encounter the disconnected vertices more than once. A nanotube with a small number of vertices in the radial direction therefore corresponds to an effectively larger flat system with the same defect density and from the earlier studies we know that larger lattices have higher $p_a$.  Due to the absence of an exit point along the radial axis, the only direction the particle can exit is the positive $y$-direction, which explains the independence of $p_a$ from the number of radial vertices.  To summarise our numerical results, we show a comparison of $p_a$ for the different geometries discussed above in Table\,\ref{ATP}.
\par
 \begin{table}[th!]
\caption{Numerically obtained transition points for finite sized systems}.
\label{ATP}
\begin{tabular}{|c|c|c|c| }
	\hline
 Size  &  $\quad\;$Square$\;\quad$    &  $\;$ Honeycomb $\;$ & $\;$  Nanotube $\;$ \\
	\hline \hline
$50 \times 50$       & 0.950 & 0.910       & 0.910    \\ 
$\;100 \times 100\;$   & 0.972 & 0.955      & 0.950  \\
$\;200 \times 200\;$   & 0.986 & 0.975       & 0.975   \\
$\;400 \times 400\;$   & 0.992 & 0.985      & 0.985    \\
\hline
\end{tabular}
\end{table}
\noindent
{\bf Analytical~:~} The differential form of the discrete-time evolution in the continuum limit allows to analytically derive an expression for the percolation probability, $\zeta(p)$, for a square lattice. It shows the clear dependence on the fraction of connections, $p$, between vertices and the lattice size $n \times n$ 
\bea
\zeta(p) \approx  p^{2n}\;,
\label{Eq:Analytical}
\eea
however it is noteworthy that it is independent of $\theta$.  The details for deriving this expression are given below in  {\bf Methods} and in Table\,\ref{ANC} we show the comparison of the transition point obtained from Eq.~\eqref{Eq:Analytical} and from the numerical analysis. Both approaches can be seen to give similar values.
 \begin{table}[th!]
\caption{Transition points for the square lattice obtained numerically and analytically.}
\label{ANC}
\begin{tabular}{|c|c|c| }
	\hline
 Size  &  $\quad\;$Numerical$\;\quad$    &  $\;$ Analytical $\;$ \\
	\hline \hline
$50 \times 50$       & 0.950 & 0.955           \\ 
$\;100 \times 100\;$   & 0.972 & 0.977       \\
$\;200 \times 200\;$   & 0.986 & 0.988          \\
$\;400 \times 400\;$   & 0.992 & 0.994         \\
$\;600 \times 600\;$   & 0.996 & 0.996         \\
\hline
\end{tabular}
\end{table}


\vskip 0.3in
\bc
{\bf \Large Discussion}
\ec

Above, we have investigated quantum percolation using a directed two-state DQW as a model for quantum transport processes. One of the main findings is that the transition point $p_a$, beyond which quantum transport can be seen, is much larger than the classical percolation threshold, $p_c$, due to localisation effects stemming from the dynamics relating to missing edges in the lattice. Therefore a small number of disconnected vertices in a large system can obstruct directed quantum transport significantly, which shows that even for the case of directed discrete-time evolution, just like for the case of continuous-time Hamiltonians\,\cite{AALR79, LN09}, Anderson localization can be a dominant process. In addition, for finite lattice sizes and unlike the classical case, we have found that $p_a$ scales with the size of the lattice tending towards unity for large lattices. This can be understood by realising that the size of the Anderson localization length in a 2D system with disorder can be quite large \cite{AALR79}, which means that on a smaller lattice one can always find a non-zero percolation probability to reach the edge of the lattice. However, this will decrease with increasing in lattice size. In our study the localization length in the $x$ direction is very small, which results in a negligible (though non-zero) contribution to the percolation probability in this direction. The directed nature of the evolution in the positive $y$ direction, however, contributes to an extended Anderson localization length, which in turn results in a non-zero percolation probability for small lattice sizes and small disorder. This can also be clearly seen in the analytical expressions derived for respective percolations along the $x$ and $y$ directions (see {\bf Methods} below).  Comparing different lattice geometries we have shown that $p_a$ is smaller for honeycomb structures and nanotube geometries  than for a square lattice of the same size.  This variation suggests that one can explore the dynamics on different lattice structures to find the one most suitable for a required purpose. For example, a system with a high $p_a$ can be well suited for quantum storage applications, whereas one with a low value will allow for more efficient transport.
 
Using two-state D-DQWs to model quantum percolation can be seen as a realistic approach to studying transport processes in various directed physical systems such as photon dynamics in waveguides with disconnected paths or quantum currents on nanotubes.  We have demonstrated its generality by allowing the parameter $\theta$ to vary randomly at each vertex and shown that this does not lead to any significant change in $p_a$. This is also confirmed by the analytical expression being independent of $\theta$. Finally, our model can be extended to two-state quantum walks on three-dimensional lattices by alternating the evolution along the different dimensions\,\cite{Cha13}. Therefore, the D-DQW  can be defined on 3D systems and a similar studies can be carried out when advanced computation resources are available.

Given the current experimental interest and advances in implementing quantum walks in various physical systems\,\cite{Do05, SSV12, SMS09, ZKG10, PLP08, PLM10, SCP10, BFL10, KFC09}, we believe that our discrete model is a strong candidate for upcoming experimental studies and its continuous form, which is detailed in the {\bf Methods}, will also be of interest for further theoretical analysis with different evolution schemes. 

Finally, it is interesting and important to understand which universality class the presented D-DQW model belongs to. While it is well known that the two-state quantum walk with disorder given by a coin parameter ($\theta$) is classified in the chiral symmetry class\,\cite{OK11}, the presence of the disconnected vertices  represent another source of disorder, which might have the potential to lead to the system moving into a different universality class. To determine this, one has to realise that the disconnected vertices lead to 8 different forms of unitary operators (see {\bf Methods}), which leads to 8 different forms for the effective Hamiltonian. Finding the universality class for our model has therefore to take the combination of the different dynamics stemming from all effective Hamiltonians into account, which is a task beyond the scope of the current work and which we will address in a future communication.
 
\begin{figure}[th!] 
\includegraphics[width=13.0cm]{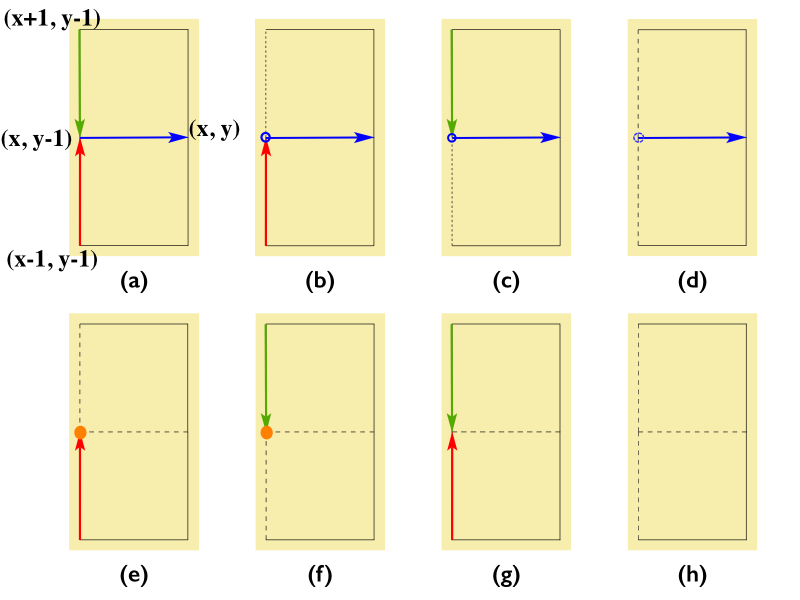} 
\caption{\footnotesize{{\bf Schematic view of the eight possible configurations encountered by a quantum state walking on a lattice with missing edges (dashed lines).} Green, red and blue arrows represent the directions of $|\downarrow \rangle$, $|\uparrow \rangle$ and both the states, respectively. 
The first row shows the four possible configurations which allow transport to the vertex $(x, y)$.  Transport  from (a) $(x+1, y-1)$ and $(x-1, y-1)$ to {\bf $(x, y)$}, (b) $(x-1, y-1)$ and $(x, y-1)$ to $(x, y)$, (c) $(x+1, y-1)$ and $(x, y-1)$ to $(x, y)$ and (d)  $(x, y-1)$ to $(x, y)$. Panel (e) shows a state being transported from $(x-1, y-1)$ getting trapped at $(x, y-1)$ and in panel (f) a state travelling from  $(x+1, y-1)$ gets trapped at $(x, y-1)$.  (g) Due to the absence of the edge from $(x, y-1)$ to $(x, y)$ the states from $(x, y-1)$ move back to $(x \pm 1, y-1)$ at the next time step and can continue to evolve in the positive $y$ direction. Panel (h) shows the absence of any transport or trapping. }  \label{fig:8}}
\end{figure}

\vskip 0.3in
\bc
{\bf \Large Method}
\ec
\noindent
{\bf Derivation of Quantum Percolation Probability on square lattice~:~}
Here we will derive the continuous limit of the percolation probability of a  two-state particle on a square lattice using the dynamics described in Fig.\,\ref{fig:2}(b), where  $|\downarrow \rangle$ and $|\uparrow \rangle$ are the basis state for evolution in $x$ direction and $|+\rangle = \alpha_{x, y}|0\rangle + \beta_{x, y}|1\rangle$ is the basis state for evolution in the $y$ direction.
To do so we need to consider all possible 8 configurations the walking particle can encounter on a broken lattice, which are schematically shown in Fig.\,\ref{fig:8}.  The first row depicts the four possible vertex configurations that result in transport along the $y$-direction, i.e.~from $(x\pm 1, y-1)$ and $(x, y-1)$  to $(x, y)$ and the second row shows the four configurations that result in transport  of the state from $(x\pm 1, y-1)$ to $(x, y-1)$ and the configuration that is trapped at position $(x, y-1)$. These latter four ones are equivalent to the configurations that result in transport of state from $(x\pm 1, y)$ to $(x, y)$ and the configuration that is trapped at position $(x,y)$. 

In the following, by approximating the unit displacement with a differential operator form,  we will derive a continuous expression describing the dynamics for all possible configurations.  By summing up the differential operator forms for each configuration, weighted by their respective probabilities, we then obtain an effective differential equation, which we use to calculate the dispersion relation and percolation probability.

\bc
\ec

\vskip 0.15in
\noindent
{\bf Figure 8(a)~:~}
For a completely connected lattice, it is straightforward to write the state of the particle at position $(x, y)$, $|\psi_{x, y}\rangle =\psi^{\downarrow}_{x, y} + \psi^{\uparrow}_{x, y}$, as function of $\theta$  in its iterative form 
\begin{align}
 \psi^\downarrow_{x,y}&=\cos(\theta)\psi^\downarrow_{x+1, y-1}-i\sin(\theta)\psi^\uparrow_{x-1, y-1 }\;, \\
 \psi^\uparrow_{x,y}&=\cos(\theta)\psi^\uparrow_{x-1,y-1}-i\sin(\theta)\psi^\downarrow_{x+1, y-1}\; .
\end{align}
These equations can be easily decoupled
\be
 \label{eq:4}
 \psi^{\downarrow (\uparrow)}_{x, y+1} + \psi^{\downarrow(\uparrow)}_{x, y-1} \;,
 =\cos(\theta) \left[\psi^{\downarrow(\uparrow)}_{x-1,y}+\psi^{\downarrow(\uparrow)}_{x+1,y}\right]\;,
\ee 
and subtracting $2\left [ 1 + \cos(\theta)\right ]\psi^{\downarrow(\uparrow)}_{x, y}$ from both sides of Eq.~\eqref{eq:4}, allows to obtain a difference form, which can be written as a second order differential wave equation
\bea
\label{KGe}
\left[\frac{\partial^2}{\partial y^2}-\cos(\theta)\frac{\partial^2}{\partial x^2} 
      + 2[1 - \cos(\theta)] \right]\psi^{\downarrow(\uparrow)}_{x, y} = 0.
\eea
Note that this is in the form of the Klein-Gordon equation\,\cite{CBS10}.

\vskip 0.15in
\noindent
{\bf Figure 8(b)~:~}
For the configuration with a missing edge from the vertex $(x, y-1)$ to $(x+1, y-1)$, the states arriving from $(x, y-2)$ and $(x-1, y-1)$ will both proceed fully to $(x, y)$, which gives
\bea
\psi_{x, y}^{\downarrow}  =&  \cos(\theta) \psi_{x, y-1} ^{\downarrow} - i\sin(\theta) \Big[\psi_{x-1, y-1}^{\uparrow} +  \psi_{x, y-1} ^{\uparrow}\Big] \;,\\  
\psi_{x, y}^{\uparrow}  =& -i  \sin(\theta) \psi_{x, y-1} ^{\downarrow} +  \cos(\theta) \Big[\psi_{x-1, y-1}^{\uparrow} +  \psi_{x, y-1} ^{\uparrow}\Big].~~
\eea
After decoupling we get
\bea
\psi_{x, y+1}^{\uparrow (\downarrow)} + \psi_{x, y-1}^{\uparrow (\downarrow)}  -\cos(\theta) \psi_{x-1, y}^{\uparrow (\downarrow)} + \psi_{x-1, y-1}^{\uparrow (\downarrow)} 
=  2 \cos(\theta) \psi_{x, y}^{\uparrow (\downarrow)}\;,
\eea
and subtracting $[2+\cos(\theta)] \psi^{\uparrow (\downarrow)}_{x, y} + \psi^{\uparrow (\downarrow)}_{x+1, y}$ on both sides lets us obtain the difference form  
\bea
\Bigg[\frac{\partial^2}{\partial y^2} + \cos(\theta)\frac{\partial}{\partial x}+ [2-3\cos(\theta)] \Bigg ]  \psi_{x, y}^{\uparrow (\downarrow)} 
 =    \Bigg[ \frac{\partial}{\partial y}  - 1 \Bigg] \psi_{x-1, y}^{\uparrow (\downarrow)}.
\eea
The right hand side can be further simplified to give
\bea
\Bigg[\frac{\partial^2}{\partial y^2} + \frac{\partial^2}{\partial y \partial x}+ [1- \cos(\theta)] \frac{\partial}{\partial x} - \frac{\partial}{\partial y} 
 + 3[1-\cos(\theta)] \Bigg ]  \psi_{x, y}^{\uparrow (\downarrow)}  = 0\;,
\eea
and the probability for this configuration is $p^2(1-p)$. 

\vskip 0.15in
\noindent
{\bf Figure 8(c)~:~}
 For configuration with a missing edge from the vertex  $(x, y-1)$ to $(x-1, y-1)$, the states arriving from $(x, y-2)$ and $(x+1, y-1)$ will both proceed fully to $(x, y)$, which gives
\bea
\psi_{x, y}^{\downarrow}  =&  \cos(\theta)\Big[ \psi_{x, y-1}^{\downarrow} + \psi_{x+1, y-1}^{\downarrow} \Big ] - i \sin(\theta) \psi_{x, y-1}^{\uparrow}\;,\\  
\psi_{x, y}^{\uparrow}  =&  -i \sin(\theta)\Big[ \psi_{x, y-1}^{\downarrow} + \psi_{x+1, y-1}^{\downarrow} \Big ] + \cos(\theta) \psi_{x, y-1}^{\uparrow}.~~
\eea
After decoupling we get 
\bea
\psi_{x, y+1}^{\uparrow (\downarrow)} + \psi_{x, y-1}^{\uparrow (\downarrow)}  -\cos(\theta) \psi_{x+1, y}^{\uparrow (\downarrow)} + \psi_{x+1, y-1}^{\uparrow (\downarrow)} 
=  2 \cos(\theta) \psi_{x, y}^{\uparrow (\downarrow)}\;,
\eea
and subtracting  $[2+\cos(\theta)] \psi^{\uparrow (\downarrow)}_{x, y} + \psi^{\uparrow (\downarrow)}_{x-1, y}$ from both sides we obtain the difference form
\bea
\Bigg[\frac{\partial^2}{\partial y^2} - \cos(\theta)\frac{\partial}{\partial x}  + [2-3\cos(\theta)] \Bigg ]  \psi_{x, y}^{\uparrow (\downarrow)}  
=    \Bigg[ \frac{\partial}{\partial y}  - 1 \Bigg] \psi_{x+1, y}^{\uparrow (\downarrow)}.
\eea
The right hand side can be further simplified to obtain
\bea
\Bigg[\frac{\partial^2}{\partial y^2} - \frac{\partial^2}{\partial y \partial x} + [1- \cos(\theta)] \frac{\partial}{\partial x}- \frac{\partial}{\partial y}  
 + 3[1-\cos(\theta)] \Bigg ]  \psi_{x, y}^{\uparrow (\downarrow)}  = 0\;,
\eea
and the probability for this configuration is $p^2(1-p)$.

\vskip 0.15in
\noindent
{\bf Figure 8(d)~:~} 
For a configuration with missing edges from the vertex $(x+1, y-1)$ to $(x, y-1)$ and from the vertex $(x-1, y-1)$ to $(x, y-1)$, the state at vertex $(x, y)$ will be,
\bea
\psi_{x, y}^{\downarrow}  =&  \cos(\theta) \psi_{x, y-1}^{\downarrow} - i \sin(\theta) \psi_{x, y-1} ^{\uparrow} \;,\\  
\psi_{x, y}^{\uparrow}  =& - i \sin(\theta) \psi_{x, y-1}^{\downarrow} + \cos(\theta) \psi_{x, y-1} ^{\uparrow},
\eea
which, after decoupling, gives
\bea
\psi_{x, y+1} ^{\uparrow (\downarrow)} + \psi_{x, y-1} ^{\uparrow (\downarrow)} = 2 \cos(\theta) \psi_{x,y}^{\uparrow(\downarrow)}\;.
\eea
Subtracting $2\psi_{x, y}^{\uparrow(\downarrow)}$ from both sides, the difference form can be obtained as
\bea
\left[\frac{\partial^2}{\partial y^2} + 2 \left [ 1 - \cos(\theta)\right ] \right]\psi^{\uparrow (\downarrow)}_{x, y} = 0\;,
\eea
and the probability for this possibility is $p(1-p)^2$. 

\vskip 0.2in
\noindent
The common feature of the next four configurations, Figs.~8(e)-(h), is the missing edge from the vertex $(x, y-1)$ to the vertex $(x, y)$, which will result in the absence of any transport along the $y$ direction. 
\vskip 0.2in
\noindent
\vskip 0.15in
{\bf Figure 8(e)~:~} 
For a configuration with missing edges from the vertex $(x+1, y-1)$ to $(x, y)$  and from vertex $(x, y-1)$ to $(x, y)$ the state will be
\bea
\psi_{x, y-1}^{\downarrow}  =&  \cos(\theta) \psi_{x, y-1} ^{\downarrow} - i\sin(\theta) \Big[\psi_{x-1, y-1}^{\uparrow} +  \psi_{x, y-1} ^{\uparrow}\Big] \;,~~\\  
\psi_{x, y-1}^{\uparrow}  =& -i  \sin(\theta) \psi_{x, y-1} ^{\downarrow} +  \cos(\theta) \Big[\psi_{x-1, y-1}^{\uparrow} +  \psi_{x, y-1} ^{\uparrow}\Big],~~
\eea
which is equivalent to
\bea
\psi_{x, y}^{\downarrow}  =&  \cos(\theta) \psi_{x, y} ^{\downarrow} - i\sin(\theta) \Big[\psi_{x-1, y}^{\uparrow} +  \psi_{x, y} ^{\uparrow}\Big] \;,\\  
\psi_{x, y}^{\uparrow}  =& -i  \sin(\theta) \psi_{x, y} ^{\downarrow} +  \cos(\theta) \Big[\psi_{x-1, y}^{\uparrow} +  \psi_{x, y} ^{\uparrow}\Big] .
\eea
After decoupling we get
\bea
2[1-\cos(\theta)] \psi_{x, y}^{\uparrow (\downarrow)} = [\cos(\theta) -1] \psi_{x-1, y}^{\uparrow (\downarrow)}\;,
\eea
and subtracting $[\cos(\theta)-1]\psi_{x,y}^{\uparrow (\downarrow)}$ from both sides we find the difference form 
\bea
[1- \cos(\theta) ] \Big [ 3 -  \frac{\partial}{\partial x} \Big ] \psi_{x, y}^{\uparrow (\downarrow)} = 0\;.
\eea
The probability for this possibility is $p(1-p)^2$.

\vskip 0.15in
\noindent
{\bf Figure 8(f)~:~} 
For a configuration with missing edges from the vertex $(x-1, y-1)$ to $(x, y)$  and from vertex $(x, y-1)$ to $(x, y)$ the state will be
\bea
\psi_{x, y-1}^{\downarrow}  =&  \cos(\theta)\Big[ \psi_{x, y-1}^{\downarrow} + \psi_{x+1, y-1}^{\downarrow} \Big ] - i \sin(\theta) \psi_{x, y-1}^{\uparrow}\;,\\  
\psi_{x, y-1}^{\uparrow}  =& - i \sin(\theta)\Big[ \psi_{x, y-1}^{\downarrow} + \psi_{x+1, y-1}^{\downarrow} \Big ] + \cos(\theta) \psi_{x, y-1}^{\uparrow}\;,  
\eea
which is equivalent to
\bea
\psi_{x, y}^{\downarrow}  =&  \cos(\theta)\Big[ \psi_{x, y}^{\downarrow} + \psi_{x+1, y}^{\downarrow} \Big ] - i \sin(\theta) \psi_{x, y}^{\uparrow}\;,\\  
\psi_{x, y}^{\uparrow}  =&  - i \sin(\theta)\Big[ \psi_{x, y}^{\downarrow} + \psi_{x+1, y}^{\downarrow} \Big ] + \cos(\theta) \psi_{x, y}^{\uparrow}\;.
\eea
After decoupling we get
\bea
2[1-\cos(\theta)] \psi_{x, y}^{\uparrow (\downarrow)} = [\cos(\theta) -1] \psi_{x+1, y}^{\uparrow (\downarrow)}.
\eea
and subtracting  $[\cos(\theta)-1]\psi_{x,y}^{\uparrow (\downarrow)}$ from both sides the difference form can be found as
\bea
[1- \cos(\theta) ] \Big [ 3 +  \frac{\partial}{\partial x} \Big ] \psi_{x, y}^{\uparrow (\downarrow)} = 0.
\eea
The probability for this possibility is $p(1-p)^2$. 

\vskip 0.15in
\noindent
{\bf Figure 8(g)~:~} 
For the configuration where only the edge from vertex $(x, y-1)$ to $(x, y)$ is missing, the state will be
\bea
 \psi^\downarrow_{x,y-1}=&\cos(\theta)\psi^\downarrow_{x+1, y-1}-i\sin(\theta)\psi^\uparrow_{x-1, y-1 }\;,\\
 \psi^\uparrow_{x,y-1}=&\cos(\theta)\psi^\uparrow_{x-1,y-1}-i\sin(\theta)\psi^\downarrow_{x+1, y-1}\; ,
\eea
which is equivalent to
\bea
 \psi^\downarrow_{x,y}=&\cos(\theta)\psi^\downarrow_{x+1, y}-i\sin(\theta)\psi^\uparrow_{x-1, y }\;, \\
 \psi^\uparrow_{x,y}=&\cos(\theta)\psi^\uparrow_{x-1,y}-i\sin(\theta)\psi^\downarrow_{x+1, y}\; .
\eea
After decoupling we get
\bea
2\psi_{x, y}^{\uparrow(\downarrow)} = \cos(\theta) \Big [ \psi_{x-1, y} ^{\uparrow (\downarrow)}+ \psi_{x+1, y}^{\uparrow (\downarrow)} \Big ]\;,
\eea
and subtracting $2\cos(\theta)\psi_{x, y}^{\uparrow (\downarrow)}$ from both sides we obtain the difference form
\bea
\Big[ \cos(\theta)\frac{\partial^2}{\partial x^2} - 2[\cos(\theta) -1] \Big ] \psi_{x, y}^{\uparrow (\downarrow)} = 0.
\eea
The probability for this possibility is $p^2(1-p)$. This situation does not lead to localisation, since the states will move back to $(x, \pm 1, y-1)$ at the next time step and can then continue to evolve in the positive $y$ direction.

\vskip 0.15in
\noindent
{\bf Figure 8(h)~:~} 
For the configuration with the three missing edges  from $(x+1, y-1)$ and $(x-1, y-1)$ to $(x, y-1)$ and from $(x, y-1)$ to $(x, y)$ all transport is suppressed and the state can be written as,
\bea
\psi_{x, y-1}^{\downarrow}  =&  \cos(\theta) \psi_{x, y-1}^{\downarrow} - i \sin(\theta) \psi_{x, y-1} ^{\uparrow}\;,\\  
\psi_{x, y-1}^{\uparrow}  = &    - i \sin(\theta) \psi_{x, y-1}^{\downarrow} + \cos(\theta) \psi_{x, y-1} ^{\uparrow} \;,
\eea
which is equivalent to
\bea
\psi_{x, y}^{\downarrow}  =& \cos(\theta) \psi_{x, y}^{\downarrow} - i \sin(\theta) \psi_{x, y} ^{\uparrow}\;,\\  
\psi_{x, y}^{\uparrow}  = &   - i \sin(\theta) \psi_{x, y}^{\downarrow} + \cos(\theta) \psi_{x, y} ^{\uparrow} . 
\eea
These expressions can be decoupled and written as
\bea
2[1-\cos(\theta)]\psi_{x, y}^{\uparrow (\downarrow)} = 0\;,
\eea
and the probability for this possibility is $(1-p)^3$.

\vskip 0.2in
\noindent
{\bf Effective differential expression~:}
Adding the differential expression for all eight case above weighted by their respective probabilities, we get 
\bea
\Bigg[\Big[p^3 + 2p^2(1-p)+ p(1-p)^2 \Big] \frac{\partial^2}{\partial y^2}  
- \Big [p^3 - p^2(1-p) \Big ] \cos(\theta) \frac{\partial^2}{\partial x^2}  
+ 2p^2(1-p) \Big [ [1-\cos(\theta)]  \frac{\partial}{\partial x} - \frac{\partial}{\partial y} \Big ] \nonumber \\
+ [1-\cos(\theta)] \Big[2p^3 + 8p^2(1-p) 
+ 8p(1-p)^2 + 2(1-p)^3 \Big ] \Bigg ] \psi_{x,y}^{\uparrow (\downarrow)}= 0 \;,
\eea 
which can be simplified as
\bea
\label{pde1}
\Bigg[ p \frac{\partial^2}{\partial y^2}  - p^2(2p-1)  \cos(\theta) \frac{\partial^2}{\partial x^2} 
+ 2p^2(1-p) \Big [ [1-\cos(\theta)]  \frac{\partial}{\partial x} - \frac{\partial}{\partial y} \Big ]  
+2 [1-\cos(\theta)] (1+p-p^2) \Bigg ] \psi_{x,y}^{\uparrow (\downarrow)}= 0.
\eea 
One can now seek a Fourier-mode wave like solution of the form 
\be
\label{eq:sol}
\psi_{x, y}^{\uparrow (\downarrow)} =  e^{i(k_x x - \omega_x y)},
\ee
where $\omega_x$ is the frequency and $k_x$  the wavenumber, which is bounded by $[0,  \sqrt{2}]$ for this differential form of the DQW\,\cite{Cha12}. 
By substituting the first and second derivative of  $\psi_{x, y}^{\uparrow (\downarrow)}$ into the Eq.\,(\ref{pde1})  we get a dispersion relation of the form, 
\bea
f(\omega_x)= p\omega_x^2 -  2p^2(1-p) i \omega_x - p^2(2p-1)\cos(\theta) k_x^2  \nonumber \\ - 2 \Big [  p^2(1-p) i k_x  + (1+p-p^2)   \Big ] [1-\cos(\theta)]  =0\;,
\eea
which can be decomposed into its real and imaginary parts to give the two conditions 
\begin{align}
\mbox{Re} [f(\omega_x)] &= p\omega_x^2 - p^2(2p-1)\cos(\theta) k_x^2  - 2 (1+p-p^2) [1-\cos(\theta)]  =0\;,\\
\operatorname{Im} [f(\omega_x) ]&= 2p^2(1-p)  \omega_x  + 2  p^2(1-p)  k_x   [1-\cos(\theta)]  =0\;.
\end{align}
The solution to the real part of the dispersion relation describes the propagation properties of the wave, whereas the  complex part relates to absorption (the trapped part in our case)\,\cite{JDJ}.  Therefore, in order to calculate the percolation probability of the propagating component, we only need to consider the solution for $\omega_x$ corresponding to the real part of the dispersion relation. This can be found as 
\bea
\omega_x(k_x)  =  \pm \sqrt{p(2p-1) \cos(\theta)k_x^2 + 2(p^{-1}-p+1)[1-\cos(\theta)]}.
\eea
\bc
\begin{figure}[th!]
\includegraphics[width=13.0cm]{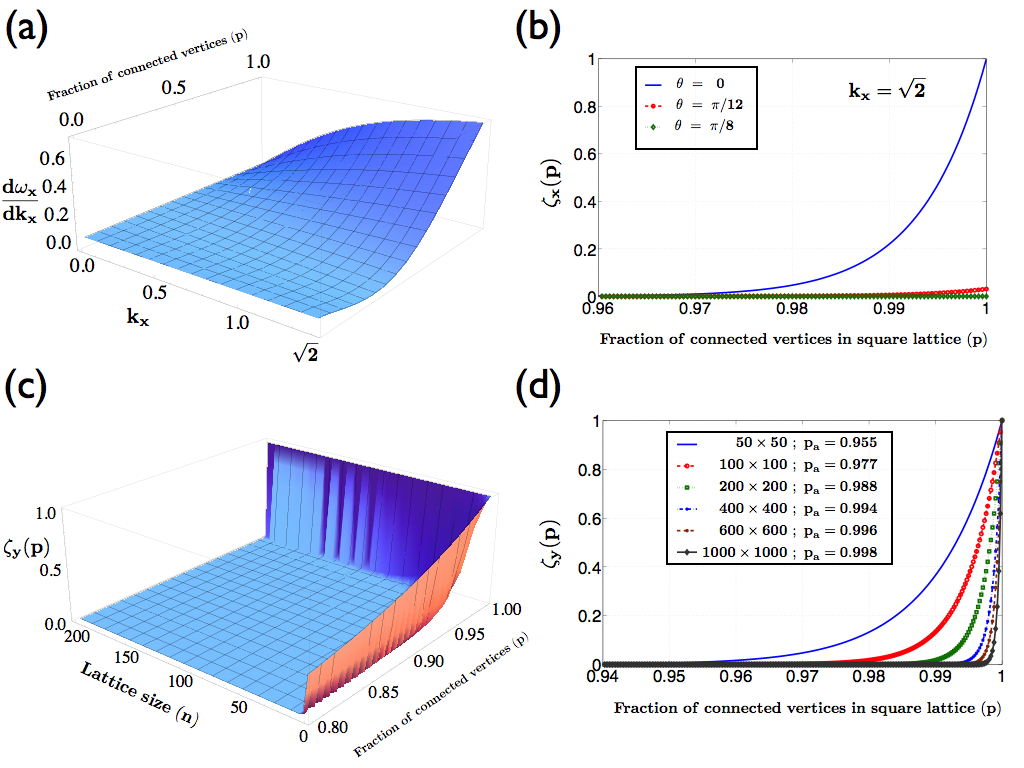} 
\caption{\footnotesize{{\bf Fraction of amplitude transport during each step and percolation probability from continuum approximation.} (a) Fraction of the amplitude transported in the $\pm x$ direction for each shift in $y$ as a function of $k_x$ and $p$ when $\theta = \pi/4$.  (b) Percolation probability along the $x-$axis for a square lattice of $50 \times 50$  for different values of $\theta$. (c) Percolation probability in the $y$ direction, $\zeta_y(p)$, as a function of lattice size $n$ and fraction of connection $p$. (d) $\zeta_y(p)$ as a function of $p$ for different lattice sizes. The obtained  transition points are identical to the values found for the discrete evolution presented in Fig.\,\ref{fig:5}.} \label{fig:9}}
\end{figure} 
\ec

The derivative of $\omega_x(k_x)$ with respect to $k_x$ describes the fraction of the amplitude $\psi_{x, y}^{\uparrow (\downarrow)}$ transported in $\pm x$ direction for each shift in the $y$ direction and in Fig.\,\ref{fig:9}(a) one can see that it increases with larger values of $k_x$ and $p$ (for $\theta=\frac{\pi}{4}$) and reaches a maximum for $k_x=\sqrt{2}$ and $p=1$. Since the transition probability will be the square of this amplitude, the percolation probability $\zeta_x(p)$ along the $x$ axis for a square lattice of $n \times n$ dimensions in the continuum limit is given by 
\bea
\zeta_x(p) =\Bigg [ \left(\frac{d \omega_x(k_x)}{d k_x} \right )^n\Bigg ] ^2\equiv  \Bigg [ \frac{p(2p-1) \cos(\theta)k_x}{\sqrt{p(2p-1) \cos(\theta)k_x^2 + 2(p^{-1}-p+1)[1-\cos(\theta)]}}\Bigg ]^{2n}\;.
\eea
 For the case of maximal transport $(k=\sqrt{2})$ we show this percolation probability for $n=50$ and for different values of $\theta$ in Fig.\,\ref{fig:9}(b). One can note that already for small values of $\theta$ (see curve for $\theta=\pi/12$) the percolation probability along the $x$ direction is very small, despite the fraction of the amplitude transported at each displacement being maximal. This is consistent with our earlier observation for the discrete evolution and leads to the conclusion that the percolation for the D-DQW is mainly due to the transition along $y-$axis. 
\par
To obtain the percolation probability along $y$-direction we go back to Fig.\,\ref{fig:8}  and write down the iterative form for transport of the state from $y-1$ to $y$ for each instance of time $t$. Defining $\Theta_{y, t} = \sum_{\forall x} \psi_{x, y, t}^{\uparrow (\downarrow)}$, all the four cases in the first row lead to
\bea
\Theta_{y, t} = \Theta_{y-1, t-1}   \implies \Theta_{y, t+1} = \Theta_{y-1, t}  \implies \nonumber \\
 \frac{\partial \Theta_{y, t} }{\partial t} =  \frac{\partial \Theta_{y, t} }{\partial y},~~~~~~~~~~~~~~~~~~
\eea
and all four cases shown in second row give 
\bea
\Theta_{y, t} = \Theta_{y, t-1}  \implies \frac{\partial \Theta_{y, t} }{\partial t} =  0 .
\eea
The differential equations for all eight configuration properly weighted with their respective probabilities is then given by
\bea
\label{partialy}
\Bigg[\Big[p(1-p)^2 +2p^2(1-p)+ p^3 \Big] \Big [\frac{\partial \Theta_{y, t} }{\partial t} - \frac{\partial \Theta_{y, t} }{\partial y}  \Big ] \nonumber \\
+ \Big [2p(1-p)^2 + p^2(1-p)+ (1-p)^3\Big ]  \Bigg ] \frac{\partial \Theta_{y, t} }{\partial t} = 0.
\eea 
Again, one can seek a Fourier-mode wave like solution with $\omega_y$ as the frequency and $k_y$ as the wavenumber of the form
\be
\label{eq:sol1}
\Theta_{y, t}^{\uparrow (\downarrow)} =  e^{i(k_y y - \omega_y t )}.
\ee
which leads to
\bea
\omega_y (k_y) = \frac{k_y\Big(p(1-p)^2 +2p^2(1-p)+ p^3 \Big ) }{3p(1-p)^2 +3p^2(1-p)+ p^3 + (1-p)^3 } = k_y p\;.
\eea
The percolation probability $\zeta_y(p)$ along the $y$-axis for a square lattice of $n \times n$ dimension in continuum limit as a function of $p$ can then be found as
\bea
\zeta_y(p) = \Bigg ( \frac{d \omega_y}{d k_y} \Bigg ) ^{2n} \equiv  p^{2n},
\eea
and we show this quantity in Fig.\,\ref{fig:9}(c)  as a function of lattice size $n$ and $p$. Since in general its value is much larger than the percolation probability in the $x$ direction we can finally write
\bea
\zeta(p) \approx  \zeta_y(p) = p^{2n}\;,
\eea
which is independent of $\theta$.  In Fig.\,\ref{fig:9}(d) we shown $\zeta(p)$ as a function of $p$ for different lattice sizes and obtain values for the transition point very close to the ones obtained numerically for the discrete evolution presented in {\bf Results}.


\vskip 0.3in
\noindent
{\bf Acknowledgements:} \\
 This work is funded through OIST Graduate University.\\
\vskip 0.1in
\noindent
{\bf Author Contributions:} \\
 CMC designed the study, carried out the numerical analysis, analytical derivation and prepared figures with support from TB.  CMC and TB together interpreted the results, and wrote the manuscript. \\
\vskip 0.1in
\noindent
{\bf Competing financial interests :}\\
The author declare no competing financial interests.

\end{document}